\documentclass{article}
\usepackage{spconf,amsmath,graphicx,hyperref}
\usepackage{booktabs}
\usepackage[table]{xcolor} 
\usepackage{enumitem}

\title{
Identity leakage through accent cues in voice anonymisation}
\name{Rayane Bakari$^{1,2}$, Olivier Le Blouch$^{1}$, Nicolas Gengembre$^{1}$, Nicholas Evans$^{2}$, Michele Panariello$^{2}$}
\address{
    $^{1}$ Orange Innovation, France \\
    $^{2}$ EURECOM, Sophia Antipolis, France \\
    \{mohamedrayane.bakari, olivier.leblouch, nicolas.gengembre\}@orange.com, evans@eurecom.fr
}

\begin{document}
\ninept
\maketitle
\begin{abstract}
Voice anonymisation is used to conceal voice identity while preserving linguistic content.
Even if anonymisation seems strong, non-timbral cues such as accent that remain post-anonymisation can help re-identification and reveal sensitive socio-demographic traits.
We report a study of residual accent information involving multiple anonymisation systems.  
We highlight the role of accent using speaker verification, accent verification, and accent classification
using a set of embeddings focusing on timbral, non-timbral and accent-related information and show the extent to which related cues facilitate reidentification post anonymisation. 
Results show that, while some systems are robust to reidentification attempts using accent cues, others leave residual, speaker-dependent, accent-related cues which can be used to reveal the voice identity. 
We also highlight accent-dependent variation in anonymisation performance, raising fairness concerns, and show that a system with character-level conditioning can help obfuscate identity-revealing accent cues, 
reducing accent-identification accuracy by 68\% on average and improving overall anonymisation performance by 11\% relative.
\end{abstract}
\section{Introduction}
\label{sec:intro}

Voice anonymisation involves the processing of a speech signal to conceal the voice identity (privacy) while preserving linguistic content and paralinguistic attributes (utility).  The VoicePrivacy Challenge (VPC)~\cite{tomashenko2020voiceprivacy} proposed standardized evaluation protocols and metrics to guide anonymisation system development.
Privacy is typically measured using automatic speaker verification (ASV), with performance estimated via the Equal Error Rate (EER), while utility measures depend on the application, including estimates of intelligibility and the speaker's emotional state, as for the most recent VPC held in 2024~\cite{tomashenko2024voiceprivacy}.

The VoicePrivacy Attacker Challenge~\cite{vpcAttackerChallenge} was launched to study vulnerabilities to attacks against anonymisation, to stimulate progress in evaluation, and hence to improve anonymisation robustness. Attack techniques reported in~\cite{zhang2024attacking, 10889147, titanetAttack, hltcoe} show that previous approaches to evaluation result in exaggerated estimates of anonymisation performance~\cite{resultsVpcAttacker}. In this paper, we show how the results of such attack studies can be used directly to improve robustness.

Most anonymisation systems~\cite{xinyuan2024hltcoe, yao2024npu, kuzmin24_spsc, gu2024ustc} modify speaker identity through the manipulation or substitution of timbral cues~\cite{spsc_bakari}. 
However, attacks that use non-timbral cues such as prosody, rhythm, speaking style, and accent can reveal the voice identity even after anonymisation~\cite{spsc_bakari}. Recent work has also shown that context-dependent duration features compromise privacy~\cite{tomashenko25_interspeech}, leading to duration-based anonymisation strategies~\cite{franzreb25_interspeech}.
The study presented in this paper investigates the influence of a speaker's accent.  
Since accent conveys socio-demographic information relating to the speaker's regional, ethnicity, and educational profile~\cite{wells1982accents}, attackers can utilise accent-related cues that persist post-anonymisation to facilitate re-identification or speaker profiling.


The use of accent-related cues has been largely overlooked in previously reported studies, while our previous work~\cite{spsc_bakari} suggests that the leakage of non-timbral cues, including accent, is a consistent vulnerability in some state-of-the-art anonymisation techniques.  
Real-world anonymisation use case scenarios further motivate our focus; previous work~\cite{meyer2025usecasesvoiceanonymization} shows that accent can be a primary identifier in small speaker pools or high-stakes contexts. 

The role and influence of accent might have been observed previously. 
Recent work~\cite{10445935} studied the variability in anonymisation performance by categorising speakers as `sheeps', `goats', `lambs', or `wolves'~\cite{doddington} based on ASV performance and human perception. 
Lambs are more easily re-identified, suggesting that, for these speakers, there remains some particular cues post-anonymisation that contribute to identity leakage.
In our previous work~\cite{spsc_bakari}, we argue that, with anonymisation typically focusing upon the manipulation or substitution of timbral cues, those remaining post anonymisation are likely to be of non-timbral origin.  We furthermore speculated that these cues are accent-related.  
The work reported in this paper is designed to confirm this hypothesis.

We investigate the extent to which accent-related cues survive anonymisation and the role that accent plays in identity leakage.
Specifically, we examine whether current anonymisation systems effectively preserve or obscure accent information, whether certain accents are systematically more or less protected, potentially raising fairness concerns and accounting for observations in~\cite{10445935}, and the potential to improve anonymisation robustness by reducing accent leakage. Our contributions are as follows:
\begin{itemize}

    \item we demonstrate that accent-related cues that persist post anonymisation contribute to identity leakage;
    \item we propose a simple criterion to qualify the accent preservation in anonymisation systems;
    \item we show biases in anonymisation performance for speakers of particular accents;
    \item we investigate an approach to mitigate accent-related identity leakage;
    \item we advocate for fairness-aware anonymisation evaluation to promote consistent protection across accents. 
\end{itemize}

\section{Methodology}
\label{sec:methodology}
\subsection{Anonymisation systems}
\label{sub_sec:anon_sys}
To address the leakage of identity from residual accent cues, we evaluate a set of baselines and competing systems submitted to the 2024 VPC~\cite{tomashenko2024voiceprivacy}. They include the B3-B5 baselines and the T8-5, T10-2, T12-5 and T25-1 participant systems defined for the VPC attacker challenge~\cite{vpcAttackerChallenge}. 

Baseline B3 combines automatic speech recognition (ASR) and text-to-speech (TTS)~\cite{b3}. B4 uses a neural audio codec (EnCodec)~\cite{panariello2024speaker}, while B5 exploits vector quantization to better disentangle linguistic and voice features~\cite{champion2023anonymizing}. B4* is built upon B4 by replacing the EnCodec decoder with the Vocos vocoder~\cite{siuzdak2024vocos}, which is trained with character-level conditioning as described in~\cite{panariello2024preserving}.

Participant systems employ specific strategies to preserve expressiveness in anonymised speech. T8-5~\cite{xinyuan2024hltcoe} randomly selects one of two methods for each utterance: a k-nearest neighbors voice conversion approach (kNN-VC~\cite{baas2023voice}) or a B3-like ASR+TTS approach. T10-2~\cite{yao2024npu} distills content and expressiveness features within a quantized neural audio codec framework. T12-5~\cite{kuzmin24_spsc} builds upon B5 with additional pitch control. Finally, T25-1~\cite{gu2024ustc} combines vector quantization and style tokens to preserve expressiveness.

\subsection{Evaluation Protocols}
We adopt three complementary evaluation tasks: \textit{Accent identification (AID)} is performed using the GenAID classifier~\cite{accentbox} to assign directly anonymised utterances to one of 13 accent classes. The Weighted Average Recall (WAR), computed both before and after anonymisation, provides a measure of how much accent information is preserved.  The WAR is computed according to: 

\begin{equation}
\label{eq:war}
R_i = \frac{TP_i}{N_i}, \quad 
\text{WAR} = \frac{\sum_{i=1}^{13} N_i R_i}{\sum_{i=1}^{13} N_i}
\end{equation}

where \(TP_i\) and  \(N_i\) are the number of correct classifications and total utterances for accent \(i\) and where \(R_i\) is the percentage of utterances in a given accent that are correctly classified.

Theoretically, the WAR can be seen as a simple criterion to qualify the accent-removal power of an anonymisation system. 
To illustrate, different cases may result depending on anonymisation architectures, including a WAR of 100\% for a system that perfectly preserves accents.
Conversely, what would be the result of a system that perfectly conceals accents? 
To assess the quality of accent management in anonymisation systems, we propose establishing a theoretical target value of 100 divided by the number of identifiable accents (7.69\% for 13 accents). This value describes, by instance, two acceptable cases of ‘perfect’ accent anonymisation: all source accents converted either to the same accent (e.g. neutral English) or to random accents.

\textit{Speaker Verification (SV)} is the well-known protocol used to evaluate the performances of speaker identification systems but transposed to anonymisation performance using genuine (same speaker) versus impostor (different speaker) trials. It is performed on embeddings extracted using GenAID, E-VPC, or W-NT, which are used to compute the cosine similarity between utterance pairs. Positive pairs contain speech from the same speaker, while negative pairs contain speech from different speakers. 

\textit{Accent verification (AV)} performs an SV-like evaluation but with a focus on accents. As in SV, embeddings are extracted and used to compute the cosine similarity between utterance pairs. But positive pairs contain speech in \emph{the same accent}, while negative pairs contain speech in \emph{different accents}.

SV and AV performances are estimated using the EER with trials balanced across speakers and accents. 

\subsection{ Datasets and Models }
We report experiments performed using the \textit{test-unseen} split of the Common Accent dataset~\cite{zuluagagomez23_interspeech}, as defined in ~\cite{accentbox} (hereafter referred to as the COMMON ACCENT). It includes utterances in 13 English accents collected from 10 speakers per accent with 10 utterances per speaker (1,300 total). Each utterance is accent-labelled, enabling controlled construction of accent verification trials. 
For speaker verification experiments, we also use the Libri-test set as defined for the 2024 VPC~\cite{tomashenko2024voiceprivacy}.

To quantify accent leakage, we leverage an accent identification (AID) model, which aims to classify accent using speech characteristics~\cite{shi2021accentedenglishspeechrecognition}. 
In particular, we use GenAID~\cite{accentbox}\footnote {Available at \url{https://github.com/jzmzhong/GenAID}}, which is trained using adversarial learning to suppress speaker-specific information and produce speaker-agnostic accent embeddings.  
The GenAID model achieves state-of-the-art performance for unseen speakers~\cite{accentbox}.
We use three off-the-shelf models to probe accent information, all of which produce embeddings: \textbf{GenAID}, used for AID, provides accent-sensitive embeddings; \textbf{E-VPC}, a variant of the ECAPA-TDNN model~\cite{desplanques2020ecapa, spsc_bakari} which is sensitive to timbral cues;  a \textbf{W-NT} model~\cite{gengembre2024disentangling} designed to capture non-timbral cues. 
The use of embeddings produced by each model allows us to evaluate and compare accent leakage along accent-specific, timbral, and non-timbral dimensions.

\subsection{Attacker scenarios}
We consider two standard VPC attack scenarios~\cite{tomashenko2020voiceprivacy}. 
For the \emph{ignorant} (\textbf{I}) scenario, the adversary compares original enrolment utterances with anonymised trial utterances without any compensation for anonymisation. For the \emph{lazy-informed} (\textbf{L}), the adversary compensates for the use of anonymisation by anonymising the enrolment utterances (using the same anonymisation system but without knowing its exact configuration) so that comparisons are made between anonymised enrolment and anonymised trial utterances~\cite{tomashenko2024voiceprivacy}.  These scenarios are particularly informative because they test model robustness without requiring retraining on anonymised data,

\section{Results}
\label{sec:results}
We evaluate identity leakage through accent cues in anonymised speech using E-VPC, W-NT, and GenAID. Our analysis considers both objective privacy in terms of EER for speaker verification (SV) and accent verification (AV).  In addition, we estimate the persistence of accent cues, quantified through the WAR, which reflects the degree to which anonymisation systems obfuscate accent-related information.

\subsection{Speaker Verification}
SV results are shown in Table~\ref{tab:results1}. 
Results confirm that the use of E-VPC embeddings consistently yields the highest EERs for most anonymisation systems and both attack scenarios.
However, higher EERs do not necessarily indicate strong anonymisation. E-VPC embeddings capture predominantly timbral cues which are less reliable after  anonymisation.
Non-timbral information, such as prosody, rhythm, accent, and speaking style, which are generally less impacted by anonymisation, are then a more reliable source of information using which the original speaker can be re-identified.
This information is better captured using W-NT embeddings, resulting in lower EERs, indicating weaker anonymisation.
EERs for GenAID embeddings are generally lower than those for E-VPC, but higher than those for W-NT. These results show that attacks against anonymisation systems using accent cues instead of timbral cues can be more effective and that, for some systems, there is little difference in SV performance using either accent or non-timbral cues. In these cases, accent cues appear to be the dominant source of reliable speaker information that remains after anonymisation.  

Across anonymisation systems, differences in EERs for I and L attack scenarios are generally modest (1-3\%). As expected, EERs are generally lower under the L attack scenario, particularly for W-NT and GenAID embeddings. For example, T10-2 and GenAID embeddings show a notable reduction from the I to L scenario. Interestingly, some systems exhibit the opposite trend.  B5 with GenAID embeddings yields a higher EER under the L than the I scenario. 
This suggests that B5 is slightly more accent-resistant in its anonymisation space than when comparing anonymised speech with original speech. 
In other words, this system still has room for improvement in terms of anonymisation, but this is not related to accents.

\vspace{-6pt}
\begin{table}[!t]
    \small
    \setlength{\tabcolsep}{6pt}
    \caption{Speaker verification EER (\%) under Ignorant (I) and Lazy-informed (L) scenarios for the Libri-test dataset. Lower values indicate weaker anonymisation.}

    \vspace{2mm} 
    \label{tab:results1}
    \centering
    \begin{tabular}{|c||cc||cc||cc|}
        \hline
        \textbf{Model} & \multicolumn{2}{c||}{\textbf{E-VPC}} & \multicolumn{2}{c||}{\textbf{W-NT}} & \multicolumn{2}{c|}{\textbf{GenAID}}\\
        \cline{2-7}
         & I & \cellcolor{gray!25} L & I & \cellcolor{gray!25} L & I & \cellcolor{gray!25} L \\ 
        \hline \hline
        \textbf{B3} & 47.4 & \cellcolor{gray!25}45.7 & 38.2 & \cellcolor{gray!25}34.7 & 46.3 & \cellcolor{gray!25} 44.1  \\
        \hline
        \textbf{B4} & 47.8 & \cellcolor{gray!25}49.5 & 34.2 & \cellcolor{gray!25}32.0 & 44.6 & \cellcolor{gray!25} 44.2  \\
        \hline
        \textbf{B4*} & 49.1 & \cellcolor{gray!25}49.8 & 35.4 & \cellcolor{gray!25} 38.6 & 44.0 & \cellcolor{gray!25}44.4  \\
        \hline
        \textbf{B5} & 49.1 & \cellcolor{gray!25}48.7 & 42.5 & \cellcolor{gray!25} 42.0 & 46.8 & \cellcolor{gray!25}48.3  \\
        \hline
        \textbf{T8-5} & 45.5 & \cellcolor{gray!25}48.2 & 32.8 & \cellcolor{gray!25} 36.3 & 46.1 & \cellcolor{gray!25} 44.9  \\
        \hline
        \textbf{T10-2} & 36.2 & \cellcolor{gray!25}35.9 & 23.6 & \cellcolor{gray!25} 22.1 & 40.9 & \cellcolor{gray!25} 38.6   \\
        \hline
        \textbf{T12-5} & 49.1 & \cellcolor{gray!25} 51.1  & 44.4 & \cellcolor{gray!25} 43.2 & 45.5 & \cellcolor{gray!25}47.1   \\
        \hline
        \textbf{T25-1} & 48.8 & \cellcolor{gray!25} 49.5 & 44.7 & \cellcolor{gray!25}44.1 & 45.2 & \cellcolor{gray!25} 47.8  \\
        \hline
    
    \end{tabular}
\end{table}


\subsection{Accent Verification}

AV results are shown in Table~\ref{tab:results2}.  With no access to T8-5, T10-2, T12-5 and T25-1 models or codes with which to anonymise the COMMON ACCENT dataset, we show results for baseline systems only. 
With only one exception, 
EERs for B3, B4* and B5 
are higher than those for B4 under all attack scenarios, indicating stronger suppression of accent cues. 
For B4, the EER for GenAID and the L scenario is lower than that for E-VPC, reinforcing our finding that accent information persists post-anonymisation and that, by focusing on timbral cues, E-VPC gives a potentially misleading interpretation of anonymisation performance. 

Of particular interest is the substantially lower EER for B4 and GenAID under the L attack scenario than for the I scenario. 
This observation suggests that,
under the L scenario, B4 tends to map homogeneous source accent clusters to homogeneous anonymized accent clusters. In other words, speakers with similar original accent are often projected similarly in anonymisation space, making identification easier for an attacker with partial system knowledge.
This also explains why B5 leads to a lower EER in the L scenario than in the I scenario, contrary to what was observed for SV experiments.
These results demonstrate that accent cues can persist after anonymisation, particularly in the case of B4, compromising privacy and highlighting the need to account for accent features in anonymisation system design.

\begin{table}[!t]
    \small
    \setlength{\tabcolsep}{6pt}
    \caption{Accent verification EER (\%) under Ignorant (I) and Lazy-informed (L) scenarios on COMMON ACCENT. Higher values indicate stronger anonymisation. Systems not available for anonymization are excluded.}
    \vspace{2mm}
    \label{tab:results2}
    \centering
    \begin{tabular}{|c||cc||cc||cc|}
        \hline
        \textbf{Model} & \multicolumn{2}{c||}{\textbf{E-VPC}} & \multicolumn{2}{c||}{\textbf{W-NT}} & \multicolumn{2}{c|}{\textbf{GenAID }} \\
        \cline{2-7}
         & I &  L & I & L & \cellcolor{gray!25} I & \cellcolor{gray!25} L \\ 
        \hline \hline
        \textbf{B3} & 50.5 & 53.7 &  47.5 &  49.7 & \cellcolor{gray!25} 51.5 & \cellcolor{gray!25} 50.6 \\
        \hline
        \textbf{B4} & 48.7 & 49.9 & 38.7  &  40.8 & \cellcolor{gray!25}48.5 & \cellcolor{gray!25}38.8  \\
        \hline
        \textbf{B4*} & 50.6 &  53.8 & 40.5  & 44.9  & \cellcolor{gray!25} 52.1 &  \cellcolor{gray!25}43.4 \\
        \hline
        \textbf{B5} & 50.5 &  49.7 & 46.3 & 48.7 & \cellcolor{gray!25} 50.2 & \cellcolor{gray!25} 49.9  \\
        \hline
    
    \end{tabular}
\end{table}

\subsection{Accent Classification}

\begin{table*}[!ht]
    \centering
    \caption{Accent identification results for the \textbf{COMMON ACCENT} dataset before and after anonymisation. Recall as defined in~\ref{eq:war}, is reported for each accent. Lower WAR indicates better accent suppression.  B2--B5 denote anonymisation systems applied to the dataset. }
     \vspace{2mm}
    \label{tab:acc_cls}
    \begin{tabular}{l|c|cccccccccccccc}
    \toprule
    \textbf{Dataset} & \textbf{WAR} & \textbf{HK} & \textbf{SA} & \textbf{ENG} & \textbf{SCO} & \textbf{US} & \textbf{SAF} & \textbf{PH} & \textbf{MYS} & \textbf{AUS} & \textbf{IRL} & \textbf{CAN} & \textbf{SG} & \textbf{NZ} \\
    \midrule
    Original & 56.77  & 44.0 & 88.0 & 78.0 & 82.0 & 20.0 & 57.0 & 80.0 & 15.0 & 81.0 & 52.0 & 76.0 & 15.0 & 50.0 \\
    B5     &  \textbf{7.69} &  0.0 & 0.0 &  0.0 &  0.0 & 24.0 &  0.0 &  10.0 &  0.0 &  7.0 &  0.0 & 56.0 &  0.0 &  3.0 \\
    B4     &  27.85 &  16.0 & 21.0 &  62.0 &  41.0 & 46.0 &  14.0 &  34.0 &  2.0 &  47.0 &  12.0 & 53.0 &  2.0 &  14.0 \\
    \rowcolor{gray!40} B4*     &  18.39 &  3.0 & 5.0 &  25.0 &  25.0 & 33.0 &  4.0 &  39.0 &  1.0 &  42.0 &  5.0 & 46.0 &  1.0 &  10.0 \\
    B3     &  9.77 &  2.0 & 0.0 &  4.0 &  2.0 & 32.0 &  0.0 &  13.0 &  0.0 &  4.0 &  1.0 & 67.0 &  0.0 &  2.0 \\
    \bottomrule
    \end{tabular}

\bigskip
\textbf{Accent abbreviations :} \\
HK = Hong Kong, SA = South Asian, ENG = English, SCO = Scottish, US = American, SAF = Southern African, \\
PH = Filipino, MYS = Malaysian, AUS = Australian, IRL = Irish, CAN = Canadian, SG = Singaporean, NZ = New Zealand.
\end{table*}

Accent classification results  
using the GenAID classifier are presented in Table~\ref{tab:acc_cls} in terms of the WAR and individually in terms of recall, for each of the 13 accents.  The first line of results is for original speech (no anonymisation).
Key observations include:
\begin{enumerate}
    \item \textbf{Degraded accent classification}: the WAR drops from 57\% for original speech to 8\% for B5 and 28\% for B4. This finding confirms the variability in accent obfuscation for different systems. 
    Obfuscation is highest for B5 for which the WAR reaches the theoretical `perfect' value of 1/13.
    In contrast, with a WAR of 28\%, B4 leaves residual accent information.
    \item \textbf{Accent-dependent leakage}: Certain accents, such as English (ENG) or South Asian (SA), remain relatively identifiable post-anonymisation (e.g., 62\%  recall for ENG using B4, and 21\% for SA), indicating that these accents are either more resistant to anonymisation or biased in their training to generate dominant accents.  Conversely, accents such as Hong Kong (HK) and Malaysian (MYS) are effectively obfuscated, with zero recall for B5 and similarly low recall for B3.
    \item  \textbf{System-dependent patterns}: B5 consistently achieves the lowest per-accent recall across most groups, indicating robust anonymisation of accent features. B4 leaves some accents partially exposed, particularly those that are subjectively similar, such as US, ENG, AUS, and CAN, highlighting a degree of accent bias in anonymisation effectiveness. B3 shows overall strong accent suppression (10\% WAR) but exhibits slightly higher recall for US (32\%) and CAN (67\%) accents. This bias stems from the mapping by B3 and B4 of source accents toward US/CAN accents, reflecting the distribution of their training data.
\end{enumerate}
To ensure faireness, future anonymisation methods should be designed to render all accents equally difficult to identify.
While the mapping of all speech to a single neutral accent (e.g., US or ENG) could, in principle, provide strong protection, anonymisation solutions should balance privacy with fairness.

\subsection{Enhanced Accent Suppression with B4*}
To address the leakage of voice identity through accent cues, especially in the case of voice conversion-based systems such as B4,
we investigated means to improve the obfuscation of accent-related information. We propose a new system B4*, an improved version of B4, to improve accent obfuscation. 
Recall and WAR results for B4* shown in Table~\ref{tab:acc_cls} show consistently better accent obfuscation compared to B4. Although the WAR for B4* (18\%) remains far above the theoretical minimum of 7.69\%, this represents a notable improvement. This is reflected by
gains observed for all accents preserved  using B4, except for PH and AUS.
These improvements are also supported by a 5\% increase in EER (11\% relative) under the L scenario in Table~\ref{tab:results2}, confirming that B4* enhances speaker anonymisation through stronger accent obfuscation.
These results demonstrate that leveraging textual conditioning, as in B4* described in~\ref{sub_sec:anon_sys}, can significantly improve anonymisation performance by explicitly targeting accent-related information.


\section{Discussion}
\label{sec:disc}
Our experiments show that current anonymisation systems effectively suppress timbral cues, but accent, a critical non-timbral feature, remains partially exposed. This leakage is both system- and accent-dependent. For instance, B5 achieves the highest overall WAR, reducing recall to zero for most accents, whereas B4 leaves accents such as ENG, US, AUS, and CAN partially identifiable. Accents like US and CAN consistently remain more detectable across systems, while HK and MYS are effectively obfuscated, reflecting the influence of training data distributions that include mostly US-accented speakers.
These results indicate that anonymisation effectiveness is uneven and may introduce residual accent bias.

The comparison between B4 and its variant B4* highlights the impact of system design. B4* improves accent suppression by conditioning the vocoder on character-level transcriptions, enforcing more canonical pronunciation and reducing speaker-specific traits. While this design improves recall across most accent groups, the observed gains may result from a combination of vocoder replacement and character-level conditioning. Future ablations are needed to disentangle the contribution of each factor. Similarly,  speech-to-text-to-speech-based systems like B3 remove most accentual information by resynthesizing speech from text, inherently eliminating accent cues. These observations suggest that leveraging textual conditioning or transcription-based approaches is a promising strategy to mitigate residual accent information.

The persistence of accent-related cues has privacy implications. Accents resistant to anonymisation could allow attackers to re-identify speakers or infer socio-demographic traits. This is especially relevant for distinctive or minority accents, where residual accent information may reveal identity more easily. Users belonging to more detectable accent groups may perceive themselves as less protected, highlighting equity concerns in privacy guarantees. 
Our findings underscore the need for evaluations that consider both objective metrics and subjective assessments by human listeners to fully capture privacy risks.

From a system design perspective, B5 emerges as the strongest system for anonymising accent-related cues, but it fails to provide fair protection across all accents, particularly for North American varieties (e.g., US, CAN). The partial confusion observed among these closely related accents lowers direct re-identification risk but still reflects a systematic bias. In contrast, the substantial residual leakage across multiple accent groups in B4 demonstrates the importance of integrating accent-aware anonymisation techniques. 
Stronger accent suppression may come at a cost. For example, resynthesis-based approaches or character-conditioned vocoders could reduce naturalness. Quantifying this trade-off with metrics such as Word Error Rate (WER) or Mean Opinion Score (MOS) is essential to balance privacy and utility. A careful evaluation framework can help identify methods that suppress accent cues while maintaining intelligibility and user satisfaction. 
More broadly, our findings motivate fairness-aware anonymisation: systems should provide consistent privacy protection across accents to mitigate bias.

\section{Conclusions}
\label{sec:concl}
This work investigates how accent cues contribute to identity leakage in voice anonymisation, using a multi-perspective evaluation that captures accent-specific, timbral, and non-timbral information. Our experiments show that accent information persists after anonymisation in some contexts, particularly when evaluated with GenAID, which outperforms a speaker-identification-focused model (E-VPC from VPC 2024) in detecting residual speaker-specific cues. Additionally, we show that anonymisation systems rarely afford uniform protection to speakers of different accents, highlighting fairness concerns.  As a first step toward accent-aware anonymisation, we explored B4*, a technique which improves accent obfuscation by  enforcing
more canonical pronunciation, confirming that privacy can be enhanced at the accent level. Future work could explore accent-agnostic anonymisation strategies, expand evaluations to cover a broader set of sociolectal traits, and integrate accent verification as a standard metric in privacy assessments. By addressing these gaps, anonymisation systems can better balance privacy, fairness, and usability across diverse populations.

The research reported here was partly supported by the ANR-23-CE23-0018 EVA project.


\let\oldthebibliography\thebibliography
\let\endoldthebibliography\endthebibliography
\renewenvironment{thebibliography}[1]{
  \begin{oldthebibliography}{#1}
    \setlength{\itemsep}{1.5pt}  
    \setlength{\parskip}{0pt}  
    \setlength{\parsep}{0pt}   
  }{
  \end{oldthebibliography}
}
\bibliographystyle{IEEEbib}
\bibliography{refs}

\end{document}